\documentclass{icrc29}
\usepackage{graphicx,amssymb,amsmath,times}
\usepackage{subfigure}
\usepackage{textcomp}
\usepackage[verbose]{wrapfig}

\begin{document}
\title[Calibration of VERITAS Telescope 1 via Muons]{Calibration of VERITAS Telescope 1 via Muons}
\author[T. B. Humensky and the VERITAS Collaboration] {T. B. Humensky$^a$, for the VERITAS Collaboration$^b$ \\
        (a) Enrico Fermi Institute, 5640 S. Ellis Ave.,
  Chicago, IL 60637, USA \\
        (b) For full author list, see J. Holder's paper ``Status and Performance of the first VERITAS Telescope,'' these proceedings
        }
\presenter{Presenter: T. B. Humensky (humensky@uchicago.edu), \  
usa-humensky-TB-abs1-og27-poster}

\maketitle

\begin{abstract}

Cherenkov light from cosmic-ray muons is a significant source of
background for the Imaging Atmospheric Cherenkov
Technique. However, muon events are also valuable as a 
diagnostic tool because they produce distinctive ring images, and
the expected amount of Cherenkov light per unit arclength can be
accurately calculated. We report  on a comparison of real and
simulated muon events in VERITAS Telescope 1, using this comparison to
validate the detector  model and to determine the
light collection efficiency of the telescope.

\end{abstract}

\section{Introduction}
Determining the absolute energy calibration of a Cherenkov telescope requires knowledge of the signal size produced by a single photoelectron.  Muons produce sharply defined ring images in the focal plane and provide the best ``test beam'' available for evaluating the end-to-end detector performance\cite{Vacanti94}.  Here, we show that simulations of muon showers and VERITAS Telescope 1  reproduce the muon energy spectrum, maximum Cherenkov angle, and the light collection efficiency of the telescope.

VERITAS Telescope 1\cite{Vpaper} is sited at the Whipple Observatory basecamp on Mt. Hopkins, at an altitude of $1275\ \text{m}$.  The telescope has a $12\text{-m}$ diameter reflector and a $499\text{-pixel}$ photomultiplier tube (PMT) camera read out by $500\text{-MHz}$ FADCs.  The PMTs are Photonis XP2970/02.  The pixel spacing is $0.15$\textdegree.  Telescope 1 has been taking data since January, 2005\cite{ICRC05}; the 5.6 hours of data used in this analysis were taken during May 9-13, 2005 at an average zenith angle of $\sim$ 40\textdegree.  Light cones had not yet been installed when this data was taken.

\section{Simulation and Analysis of Muon Images}
Cherenkov light is produced by charged particles when their speed exceeds the speed of light in the ambient medium, $\beta > 1/n(\lambda)$.  At the altitude of Telescope 1, the threshold energy for a muon to produce Cherenkov light is $\sim 4.6\ \text{GeV}$, and the Cherenkov angle of a $20\text{-GeV}$ muon is within 0.01\textdegree\ of its maximum value of 1.32\textdegree.

Simulated muons are generated using Corsika 6.031\cite{corsika} for the air showers and GrISU\cite{grisu} for the detector simulation.  Muons are scattered with impact parameters within a radius of $20\ \text{m}$ and at angles up to 1\textdegree\ off the telescope axis.  They are injected at a distance of $2\ \text{km}$ ($1.5\ \text{km}$ vertical distance) from the telescope.  The atmosphere is modelled using a Corsika external atmosphere appropriate for midlatitude winter (discussed below) and atmospheric extinction is modelled using MODTRAN 4\cite{mkd_thesis}.  Measurements of the mirror reflectivity are used over the range $260\text{-}700\ \text{nm}$, and are extrapolated down to $200\ \text{nm}$.  The PMT quantum efficiency (QE) is taken from manufacturer specifications covering the full range $200\text{-}700\ \text{nm}$.  FADC traces which are not part of an image are extracted from the data runs and used to provide pedestal traces for the simulations to accurately reproduce fluctuations from night-sky background and electronics noise.  Individual pixel gains and timing offsets are also applied to the simulations.  A broken power law is used for the energy spectrum, with the spectral indices inferred from~\cite{bugaev}:  indices of $\text{-}1.93$ and $\text{-}2.34$ over the energy ranges $4.5\text{-}10\ \text{GeV}$ and $10\text{-}20\ \text{GeV}$, respectively.  
Equal numbers of $\mu^+$ and $\mu^-$ are simulated.

\begin{wrapfigure}[19]{r}{0.5\textwidth}
\begin{center}
\includegraphics*[width=0.5\textwidth,clip]{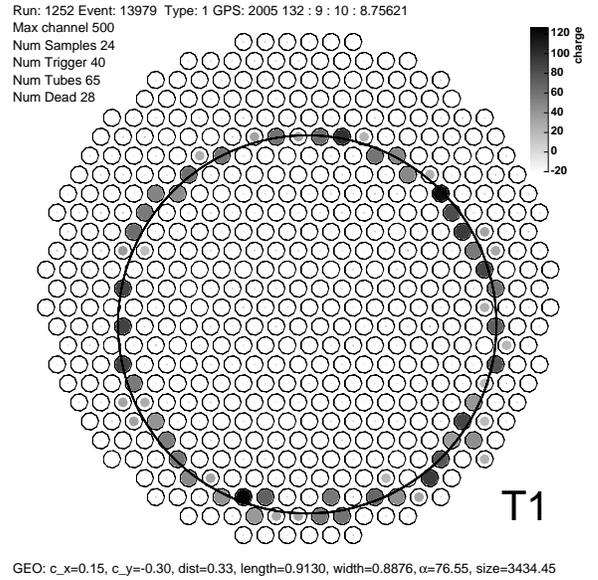}
\caption{\label {fig:muon} Typical muon image.}
\end{center}
\end{wrapfigure}

Images are characterized according to a circle fit, minimizing the function
\begin{equation}
\chi^2 = \sum_{i=1}^{N_{tubes}} w_i \centerdot (\sqrt{ (x_i - x_0)^2 + (y_i - y_0)^2 } - R)^2,
\end{equation}
where $w_i$ is the signal amplitude in pixel $i$, $x_0$, $y_0$, and $R$ are the fit parameters of the circle, and $x_i$ and $y_i$ are the coordinates of the $i^{th}$ pixel in the image.  A two-pass fitting procedure is used:  after the first pass, pixels which are more than 0.3\textdegree\ away from the circle are thrown out, and the image is fit again.  This procedure improves the fit in cases where a muon is accompanied by light from an associated hadronic shower.  Figure~\ref{fig:muon} shows a muon image recorded by the VERITAS Telescope 1 camera.  The black circle is a fit to the image.

Muon images are selected from the data files on the basis of a number of cuts: 

\begin{itemize}
\item  Hillas width\cite{Hillas85} of the image (length of the semi-minor axis) $ > 0.5\text{\textdegree}$, 
\item  image size (sum of FADC signals in pixels included in the 2nd-pass fit) $ > 3000$ digital counts (dc), 
\item  RMS timespread in the arrival of light in the pixels $ < 3.0\ \text{ns}$, 
\item  RMS width of the Cherenkov ring $ < 0.12\text{\textdegree}$, 
\item  completeness of the ring ($> 75\ \%$ of azimuthal bins with a size above a 50-dc threshold), 
\item ring center within 0.4\textdegree\ of the center of the field of view, and 
\item a negligible change in ring parameters between the first- and second-pass fits.  
\end{itemize}
These cuts select complete and nearly complete rings with very little contamination from hadronic showers.  Except where otherwise noted, all results below utilize all of the above cuts.

Following the application of cuts, the simulated and data image size distributions are compared, and a scale factor $f_S = \frac{<size>_{sim}}{<size>_{data}}$ is determined.  If necessary, a second set of simulated muons is then generated with the PMT QE scaled by $f_S$, independent of wavelength.  This process is repeated until $f_S$ converges to unity.  The net required change in QE is then a measure of our understanding of the telescope's light collection efficiency. 

\section{Results and Discussion}
\begin{figure}[h]
\begin{center}
\includegraphics*[width=\textwidth]{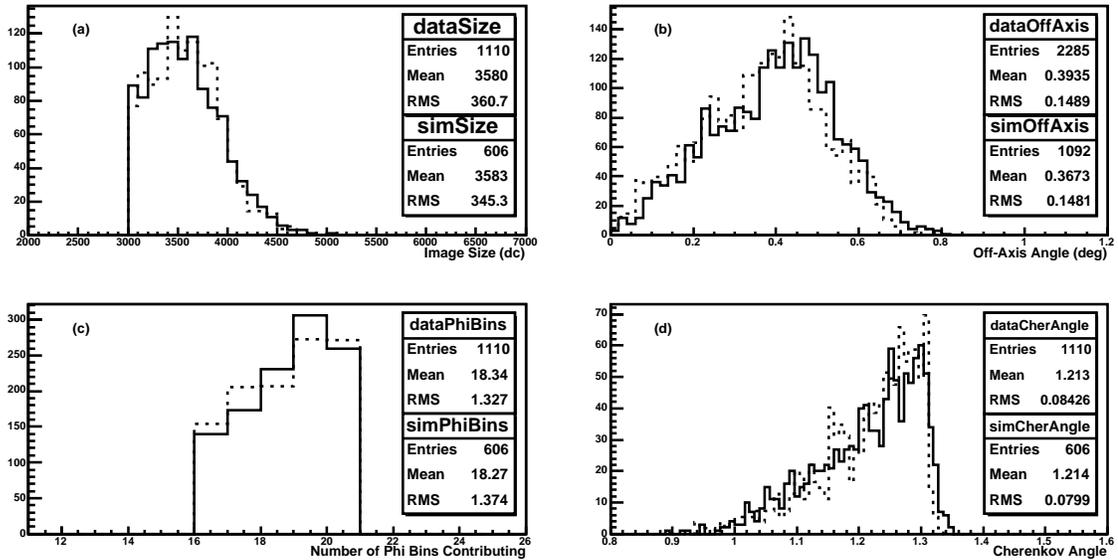}
\caption{\label {fig:finver} Properties of muon rings.  Solid line:  data.  Dotted line:  simulation. Simulated distributions have been scaled to have the same area as the data distributions.  (a) Size spectrum.  (b) Angle between muon direction and telescope axis.  (c) Azimuthal completeness of rings. (d) Cherenkov angle.}
\end{center}
\end{figure}

The results shown below required no iteration.  Figure~\ref{fig:finver} shows that the simulated muons reproduce basic characteristics of the data reasonably well.  In each case, the simulated curves have been rescaled so that the area of the simulated histogram (dotted line) equals that of the data histogram (solid line).  In addition, the simulated image sizes (Figure~\ref{fig:finver}a) have been rescaled by $f_S$ = 1.0034 so that their mean equals that of the data. The distribution of rescaled simulation image sizes agrees well with the data.  The distributions of muon angles with respect to the telescope optic axis (Figure~\ref{fig:finver}b, with cut at 0.4\textdegree\ removed) and the azimuthal distributions of light (Figure~\ref{fig:finver}c) also agree well, showing that the simulations sample the same distribution of off-axis angles and impact parameters as the data.

The shape of the Cherenkov angle distribution (Figure~\ref{fig:finver}d) depends on the muon energy spectrum, and it is likely that the muon spectrum is softer than assumed at a zenith angle of 40\textdegree.  The maximum Cherenkov angle depends on the index of refraction near the observation level; the good agreement (within $\sim 0.01\text{\textdegree})$ between simulations and data on the maximum Cherenkov angle ($\sim 1.32\text{\textdegree}$) is achieved by using an atmospheric model appropriate for midlatitude winter.  A more appropriate model, the US 1976 Standard Atmosphere, underestimated the maximum Cherenkov angle by $\sim 0.04\text{\textdegree}$.  Several possible explanations for this discrepancy are under investigation.

\begin{wrapfigure}[21]{r}{0.5\textwidth}
\begin{center}
\includegraphics*[width=0.5\textwidth,clip]{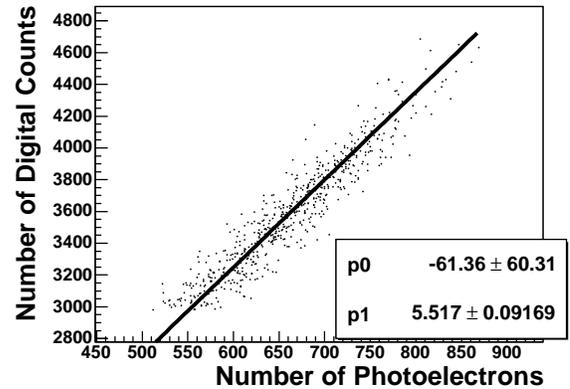}
\caption{\label {fig:dcperatio} Image size versus number of photoelectrons for simulated muon events.  The slope of the line is the dc/pe ratio, a measure of the absolute gain of the telescope.  The fit parameters define a line $y=p0+p1*x$.  Error bars have been suppressed for clarity.}
\end{center}
\end{wrapfigure}

The dc/pe ratio is calculated from simulations as in Figure~\ref{fig:dcperatio} by selecting simulated events that pass the above cuts and fitting a line to a  plot of the size in digital counts (dc) of the  ring versus the number of photoelectrons generated by the simulation.  
The dc/pe ratio is the slope of the line, $5.52 \pm 0.09\ \text{dc/pe}$, which is in good agreement with the value found via special laser runs, $5.3 \pm 10\ \%_{syst}\ \text{dc/pe}$.  It also agrees with 
the expected value of 5.57 based on the design parameters of the electronics chain, which verifies that the simulation is working as expected.

If we take the dc/pe ratio as well measured, then we can use the scale factor $f_S$ to verify our understanding of the telescope's light collection efficiency.  The telescope's light collection efficiency is determined by several 
factors:  atmospheric extinction and scattering, mirror reflectivity, PMT QE, and the packing fraction of the camera.  
For this work, it is convenient to scale the PMT QE by $f_S$.  By iteratively generating simulated muons, each time scaling the PMT QE, 
we can determine the adjustment to the collection efficiency required to achieve $f_S = 1$.  We find that the base simulation, modified to include the extrapolation of mirror reflectivity over $200\text{-}260\ \text{nm}$, reproduces the telescope's collection efficiency to within $\sim 0.5\%$ on the first iteration.  An earlier 
series of iterations which neglected this extrapolation required an efficiency adjustment of $\sim 13\ \%$, consistent with the indication from simulations that for local muons approximately $10\ \%$ of the detected light is below $260\ \text{nm}$.  Note that for gamma rays the contribution below $260\ \text{nm}$ to the detected light is negligible.

\section{Conclusion}
Comparison of muon rings in VERITAS Telescope 1 data with simulations demonstrate that the simulation of the VERITAS Telescopes reproduces well the properties of the images.  Combined with measurements of the single-PE amplitude via dedicated laser runs, this study shows that the base simulation reasonably reproduces the light-collection efficiency of the telescope.

A dedicated high-multiplicity trigger is planned which will allow the individual telescopes of the VERITAS array to acquire muon images concurrent with regular data taking for continuous monitoring of the telescopes' absolute gain and light collection efficiency.

\section{Acknowledgements}
This research is supported by grants from the U.S. Department of Energy, the National Science Foundation, the Smithsonian Institution, by NSERC in Canada, by Science Foundation Ireland, and by PPARC in the UK.

\end{document}